\documentclass{article}
\usepackage[preprint]{spconf}
\usepackage{spconf,amsmath,graphicx}
\usepackage{threeparttable}
\usepackage{multirow}
\usepackage{multicol}
\usepackage{booktabs}
\usepackage{hyperref}
\usepackage{footnote}
\usepackage{comment}
\usepackage{algorithm}
\usepackage{caption}
\usepackage[labelformat=simple]{subcaption}

\usepackage[noend]{algpseudocode}
\newcommand{\pluseq}{\mathrel{{+}{=}}}

\copyrightnotice{\copyright\ IEEE 2021}
\toappear{To appear in {\it Proc.\ ICASSP2021, June 6-11, 2021, Toronto, Canada}}

\title{A parallelizable lattice rescoring strategy \\with neural language models}
\name{Ke Li$^{1}$, Daniel Povey$^3$, Sanjeev Khudanpur$^{1,2}$
}
\address{
  $^1$Center for Language and Speech Processing \& $^2$Human Language Technology Center of Excellence \\ 
  The Johns Hopkins University, Baltimore, MD 21218, USA.\\
  $^3$Xiaomi Corp., Beijing, China.}
\email{
\texttt{\{kli26,khudanpur\}@jhu.edu}, \texttt{dpovey@gmail.com}}
\begin{document}
\ninept
\maketitle
\begin{abstract}
This paper proposes a parallel computation strategy and a posterior-based lattice expansion algorithm for efficient lattice rescoring with neural language models (LMs) for automatic speech recognition.
First, lattices from first-pass decoding are expanded by the proposed posterior-based lattice expansion algorithm. Second, each expanded lattice is converted into a minimal list of hypotheses that covers every arc. Each hypothesis is constrained to be the best path for at least one arc it includes. For each lattice, the neural LM scores of the minimal list are computed in parallel and are then integrated back to the lattice in the rescoring stage. 
Experiments on the Switchboard dataset show that the proposed rescoring strategy obtains comparable recognition performance and generates more compact lattices than a competitive baseline method. Furthermore, the parallel rescoring method offers more flexibility by simplifying the integration of PyTorch-trained neural LMs for lattice rescoring with Kaldi. 


\end{abstract}
\section{Introduction}
\label{sec:intro}
Neural language models (LMs), including long short-term memory (LSTM) and Transformer based ones, have significantly improved performance over $n$-gram LMs in automatic speech recognition (ASR)~\cite{mikolov2010recurrent,chen2015recurrent,xu2018neural,zeghidour2018fully, synnaeve2019end, irie2019language,li2020empirical}. 
Since it is challenging for one-pass decoding with a neural LM to obtain competitive performance with lower latency than a two-pass approach~\cite{hori2014real,shi2014efficient,sundermeyer2015feedforward,beck2019lstm,jorge2020lstm}, a widely adopted way is still to use neural LMs to rescore $N$-best hypotheses (alternative word-sequences) or lattices that are decoded with an $n$-gram LM~\cite{deoras2011fast,sundermeyer2014lattice,liu2014efficient,sundermeyer2015feedforward,liu2016two,chen2017future,kumar2017lattice,xu2018pruned}. 
A lattice is a compact representation of the hypothesis space for an utterance. $N$-best hypotheses only cover a small subspace. Thus, lattice rescoring usually outperforms $N$-best rescoring.

The key for lattice rescoring is to balance accuracy and efficiency since exact rescoring is not practical because it involves expanding a lattice into a linear or prefix tree structure and rescoring each hypothesis. A major speed bottleneck in lattice rescoring using a neural LM is the LM evaluation. 
Neural LM probabilities are usually computed on-the-fly and sequentially among hypotheses in a lattice during lattice traversal~\cite{sundermeyer2014lattice,liu2016two,xu2018pruned}. Though caching computed probabilities~\cite{liu2016two} or pruning-based methods~\cite{sundermeyer2014lattice,xu2018pruned} can reduce the number of evaluations, the sequential order of LM evaluation in a lattice is inefficient. The process can be accelerated significantly by evaluating multiple hypotheses in parallel. However, given the graph structure of lattices, taking advantage of such speedup is challenging, especially for lattice rescoring methods that perform expansion and rescoring simultaneously.
To enable batch computation, we convert a lattice into a minimal list of hypotheses that satisfy two conditions. First, every arc should be included in at least one hypothesis. Second, each hypothesis is the best path for at least one arc it contains. Computed neural LM scores are integrated back into the lattice for rescoring where score refer to negative log probabilities.

Lattice rescoring usually involves lattice expansion. Performing rescoring without changing the lattice structure is feasible, but it is generally not as good as with expansion~\cite{sundermeyer2014lattice}. To prevent expanded lattices from being too large, equivalence estimation of history states and pruning-based methods have been proposed~\cite{sundermeyer2014lattice,liu2014efficient,sundermeyer2015feedforward,liu2016two,chen2017future,xu2018pruned}. For example, an $n$-gram approximation method restricts lattice size by merging history states that share $(n-1)$ most recent words. But $n$-gram approximation based expansion method may sacrifice accuracy 
and waste computation on less likely paths. 
The general goal of lattice expansion is to make arcs on relatively probable paths have unique histories so that neural LM scores for them can be exact. To this end, we propose a new lattice expansion method that expands arcs only when their posteriors are larger than a threshold. Effectively, only more probable arcs are expanded so that arcs on sufficiently likely paths tend to have unique histories.

In summary, we propose an efficient lattice rescoring strategy that enables parallel computation of neural LM scores within a lattice. 
The strategy mainly involves operations such as posterior-based lattice expansion and lattice-to-list conversion using a proposed path cover algorithm. 
Furthermore, we experiment with a refined lattice rescoring strategy to further improve results. The proposed lattice-to-list conversion makes it easier to integrate neural LMs trained with PyTorch (or other tools) for efficient lattice rescoring in Kaldi~\cite{povey2011kaldi}\footnote{The code and recipes on SWBD and WSJ are available in Kaldi: \href{https://github.com/kaldi-asr/kaldi/blob/master/egs/swbd/s5c/local/pytorchnn/run_nnlm.sh}{kaldi/egs/swbd/s5c/local/pytorchnn/} , \href{https://github.com/kaldi-asr/kaldi/blob/master/egs/wsj/s5/local/pytorchnn/run_nnlm.sh}{kaldi/egs/wsj/s5/local/pytorchnn/}.}. 

\vspace{-3mm}
\section{Lattice Conversion and Expansion}
\subsection{Lattices}
A lattice is a graph representation of hypothesis space for an utterance and 
it can encode an exponential number of hypotheses with respect to the number of states. A lattice has one start state and a set of final states. A path in a lattice is consecutive transitions from the start state to a final state. Assuming lattices generated from first-pass decoding in a weighted finite state transducers (FST) based ASR system are determinized, each path represents a unique word-sequence. 

Next, we will introduce the methods for lattice-to-list conversion, estimation of neural LM scores for each arc, and the posterior-based lattice expansion. 

\subsection{Lattice-to-List Conversion}
Many lattice rescoring methods compute neural LM scores for arcs within a lattice dynamically during traversing the lattice. Considering that neural LM evaluation is a major speed bottleneck and the sequential order of traversing a lattice is inefficient, we propose a method to enable batch computation of hypotheses within a lattice. The general idea is to convert a lattice into a list of hypotheses that include every arc. Neural LM scores are then computed in parallel and merged back into the lattice. 

A lattice $L$ can be viewed as a weighted directed acyclic graph $(V, E)$, where $V$ and $E$ denote the set of states (vertices) and arcs (edges), respectively. We define a \textit{path cover} of lattice $L$ as a set of paths such that every arc in $E$ is included in at least one path in the set. A \textit{minimal path cover} of $L$ is a path cover containing fewest possible paths. Our definition of path cover is different from the original definition in graph theory in which paths should cover states rather than arcs and they may start and end anywhere. 

The size of a minimal path cover can be determined as
\begin{equation}
\sum_{s \in V}(\max(\text{deg}_\text{out}[s] - \text{deg}_\text{in}[s], 0))
\end{equation}
where $\text{deg}_\text{out}[s]$ and $\text{deg}_\text{in}[s]$ are the number of outgoing and incoming arcs of state $s$, because for each state, extra outgoing arcs should be covered by extra paths. Considering efficiency, the list of hypotheses converted from a lattice should be a minimal path cover. 

However, since neural LM scores computed from the list directly affect rescoring, the quality of the hypotheses may matter more than the size. For each arc, a common choice for its history is the one in the best path that contains the arc.
Therefore, a straightforward way of generating the list is: i) take the best path that contains each arc and sort them, ii) from the worst path to the best path, remove one if removing it does not cause any arc uncovered. While this method is not optimal since some generated paths are redundant and need to be removed. To fix it, we record the best path information for each arc during path generation so that it will not be regenerated if it already exists.
The pesudocode for this method is shown below. 

\begin{algorithm}
\caption{A Constrained Path Cover Algorithm}
\label{alg1}
\hspace*{\algorithmicindent} \textbf{Input:}  \textit{$L$}: a lattice\\
 \hspace*{\algorithmicindent} \textbf{Output:} \textit{$O$}: a list of paths, each is represented as a linear FST.
\begin{algorithmic}[1]
\Procedure{ConstrainedPathCover}{$L$}       
    \State ToplogicalSort($L$) 
    \State $P\gets$ [] \Comment{A list of pairs of a path and its cost}
    \State $\alpha, \beta\gets$ ViterbiForwardBackward($L$)
    \For{$s = 0: S-1$} \Comment{Loop over states}
    \For{$e \in s.out$} \Comment{Loop over outgoing arcs of $s$}
    \If{best path including $e$ is not generated}
    \State $p, c \gets$ BestPathForAnArc($\alpha$,$\beta$,$s$,$e$) 
    \State $P$.append(($p, c$))
    \EndIf
    \EndFor
    \EndFor
    \State Sort($P$) \Comment{Sort paths based on their costs}
    \State $O\gets$ ConstructOutputLattice($P$)
\EndProcedure
\end{algorithmic}
\end{algorithm}
\vspace{-2mm}
``Constrained'' means that each path must be the best path for at least one arc it includes. 
The ViterbiForwardBackward function in Algorithm~\ref{alg1}
computes best costs $\alpha$ and $\beta$ from the start state to every other state and from every final state to every other state, respectively. It also records the best predecessor and successor states of each state so that best paths can be found. The linear FSTs that represent the best paths are then converted into word-sequences 
for neural LM evaluation. 

\subsection{Estimation of Neural LM Scores}
Neural LM scores of word-sequences converted from a lattice by the path cover algorithm need to be integrated back into the lattice for rescoring. If an arc in the lattice is shared by multiple paths, there are multiple neural LM scores associated with the arc. An approximation thus needs to be made to assign a single neural LM score for the arc. We experiment with three ways for obtaining the approximation: (i) by simply averaging the neural LM scores from the shared paths, (ii) obtaining a refined estimation using a weighted average, where weights are normalized values of neural LM scores of histories for the arc, and (iii) choosing the neural LM score from the lowest-cost path among the shared paths. Note, the lowest-cost path is not guaranteed to be the best path including the arc in the lattice because of the way the list of word-sequences generated. We thus refer to the third estimation as ``semi-Viterbi'' in the experiment.


%


\vspace{-2mm}
\subsection{Posterior-based Lattice Expansion}
Lattice rescoring usually involves lattice expansion. To prevent the expanded lattices from blowing up in size, a commonly adopted approach is an $n$-gram approximation of history states. It merges history states with the same $(n-1)$ most recent words. However, it sacrifices accuracy since histories for computing neural LM scores may not be unique for many arcs. It also may expand out many paths with low probability, which is not optimal. To alleviate the problems of $n$-gram approximation, we propose a new expansion method. It expands arcs with posteriors higher than a threshold $\epsilon \in (0,1)$. This method aims to make arcs on relatively probable paths have unique histories so that neural LM scores can be exactly computed. We refer to this method as posterior-based lattice expansion as present below.

\vspace{-1mm}
\begin{algorithm}
\caption{A Posterior-based Lattice Expansion Algorithm}
\label{alg2}
\hspace*{\algorithmicindent} \textbf{Input:}  \textit{$L_\text{in}$}: a lattice; \textit{$\epsilon$}: a threshold for arc posteriors\\
 \hspace*{\algorithmicindent} \textbf{Output:} \textit{$L_\text{out}$}: an expanded lattice
\begin{algorithmic}[1]
\Procedure{PosteriorExpansion}{$L_\text{in}$, $\epsilon$}       
    \State ToplogicalSort($L_\text{in}$) 
    \State $\mathbf{\alpha}\gets$ [] \Comment{Initialize forward logprobs for states in $L_\text{out}$}
    \State $\beta\gets$ BackwardCosts($L_\text{in}$) \Comment{$\beta[0]$ is the total logprob}
    \State $L_\text{out}.\text{SetStart(0)}$ \Comment{Add start state $0$ in $L_\text{out}$}
    \State $M[(0,0)]\gets 0$ \Comment{Initialize state map from a state pair ($s_\text{in}, s_\text{out}$) to a state $s_\text{out}$, where $s_\text{in} \in L_\text{in}, s_\text{out} \in L_\text{out}$}
    \State $Q.$push((0, 0)) \Comment{Initialize the queue of state pairs}
    \While{$!Q$.empty()}
    \State $s_\text{in},$ $s_\text{out} \gets$ DeQueue($Q$)
    \For{$e \in s_\text{in}.out$} \Comment{Loop over outgoing arcs of $s_\text{in}$}
    \State $s_\text{in}^\text{next} \gets$ $e$.nextstate
    \State $a_{e} \gets$ $\alpha[s_\text{out}]+ e.$weight \Comment{weight is $-$logprob}
    \State $e_\text{post} \gets \text{exp}(a_{e} + \beta[s_\text{in}^\text{next}] - \beta[0])$ \Comment{arc posterior}
    \If{$e_\text{post}$ $>$ $\epsilon$}
    \State $s_\text{out}^\text{next} \gets$ $L_\text{out}.$AddState()
    \State $Q.$push(($s_\text{in}^\text{next},$ $s_\text{out}^\text{next}$))
    \State $M[(s_\text{in}^\text{next}, s_\text{out}^\text{next})] \gets s_\text{out}^\text{next}$
    \ElsIf{$s_\text{in}^\text{next}$ is never copied to $L_\text{out}$}
    \State Repeat line 15-17 and mark $s_\text{in}^\text{next}$ as copied
    \Else
    \State $s_\text{out}^\text{next} \gets$ GetCopyState($s_\text{in}^\text{next}$) 
    \EndIf
    \State $L_\text{out}.$CreateArc($M[(s_\text{in}, s_\text{out})],$ $e,$ $s_\text{out}^\text{next}$)
    \State $\alpha[s_\text{out}^\text{next}]\pluseq a_{e}$ \Comment{Update forward logprobs}
    \EndFor
    \EndWhile
\EndProcedure
\end{algorithmic}
\end{algorithm}

Algorithm~\ref{alg2} is a composition-type algorithm mainly implemented with a queue of state pairs. Each pair represents a state in the input lattice and its copy state in the expanded lattice. The basic question for lattice expansion is whether an incoming arc should be split off from the rest of the incoming arcs to its destination state. The rule is to allocate a new copy of the destination state if the arc posterior is larger than $\epsilon$, otherwise transition to the original destination state. The threshold $\epsilon$ controls the size of expanded lattices such that larger $\epsilon$ results in smaller lattices. The backward costs $\beta$ are computed in advance while the forward costs $\alpha$ are computed dynamically during creating the expanded lattice. 


\section{Lattice Rescoring Strategy}
\subsection{Non-iterative Lattice Rescoring}
\label{sec:non_iterative}
Combining the posterior-based lattice expansion algorithm and constrained path cover method, we propose an efficient lattice rescoring strategy that enables batch computation for words within a lattice. It mainly consists of two steps. First, lattices from first-pass decoding are expanded by the posterior-based lattice expansion method. We apply beam pruning before lattice expansion since in practice, we observe that it can reduce lattice size without hurting performance. 
Second, each expanded lattice is converted into a list of hypotheses. The neural LM scores that are computed in parallel are approximated when necessary and merged back into the expanded lattices for rescoring. We then find the best path in each rescored lattice and compute WERs. 
The proposed lattice rescoring strategy is referred to as ``non-iterative'' to distinguish it from a two-pass rescoring method introduced in section~\ref{sec:iterative}. 

When neural LM scores are put back to lattices, they are interpolated with LM scores of the original $n$-gram LM
The interpolation involves removing a portion of the original LM scores from the lattice, which is implemented by FST composition. 


\vspace{-2mm}
\subsection{Iterative Lattice Rescoring}
\label{sec:iterative}
To further improve the performance of the non-iterative lattice rescoring method described above, we propose a refined approach which introduces an extra rescoring stage. First, the original $n$-gram LM scores on decoded lattices are replaced with neural LM scores while the lattice structure is fixed. The proposed non-iterative rescoring strategy is then applied to the resulting lattices. We expect the integrated neural LM scores from the score replacement stage can result in better path cover lists and thus more accurate recognition results than the non-iterative method alone. 
 
We refer to the refined rescoring approach as ``iterative'' since rescoring are executed twice. An alternative way is to perform the non-iterative rescoring twice. But it complicates the rescoring procedure and slows the rescoring speed by introducing an extra lattice expansion operation.

\vspace{-2mm}
\section{Experiments}
\subsection{Datasets and Setups}
We conduct experiments on the telephone speech corpus Switchboard (SWBD) which consists of approximately 260 hours of speech. We use Kaldi for acoustic model training and decoding. The acoustic model is factorized TDNNs~\cite{povey2018semi} trained with the LF-MMI objective~\cite{povey2016purely}. The audio data of English Fisher corpus is not included. We use Kaldi RNNLM~\cite{xu2018neural} for text data preprocessing. 
There are a total of 34M words in the training dataset. 

We experiment with both LSTM and Transformer. They are word-level LMs with vocabulary around 30K and trained with PyTorch. We use a 2-layer LSTM model with hidden dimension 650, and a 6-layer Transformer model with 8 heads and 512 hidden dimension. The LSTM and Transformer LMs have a total of 26.5M and 25M parameters respectively, both with parameter tying. 
We refer the readers to~\cite{li2020neural} for further details about the models.

Besides, we train an LSTM LM with Kaldi 
to compare the proposed rescoring method with the pruned lattice rescoring algorithm~\cite{xu2018pruned}. We do not use PyTorch-trained LMs for comparison since integrating them into the pruned rescoring algorithm is relatively complicated. That is also a motivation to develop the new lattice rescoring strategy.

\subsection{Effect of Estimation Methods}
We evaluate the performance of the three approximation methods for neural LM scores with both LSTM and Transformer models using the non-iterative lattice rescoring strategy. 
WERs in Table~\ref{tab:arc_estimation} show that the semi-Viterbi estimation consistently outperforms the other two. It is thus used in all the remaining experiments.
\begin{table}[ht]
    \vspace{-3mm}
    \setlength{\tabcolsep}{2.0pt}
    \caption{WERs (\%) on Hub5'00 (full set) of SWBD from the non-iterative lattice rescoring strategy with three estimation methods.}
    \label{tab:arc_estimation}
    \centering
    \scalebox{0.9}{
    \begin{tabular}{ l  c  c  c  c }
    \toprule
         Model & $\epsilon$ &Average & Weighted Average & Semi-Viterbi \\
        \midrule
        \multirow{2}{*}{LSTM} & 0.5 & 10.8 & 10.8 & \textbf{10.7}\\
        & 0.05 & 10.7 & 10.7& \textbf{10.6} \\
        \midrule
        \multirow{2}{*}{Transformer} & 0.5 & 10.7 & 10.7& \textbf{10.6} \\
        & 0.05 & 10.6 & 10.6 &\textbf{10.5} \\
        \bottomrule
    \end{tabular}}
\end{table}
\vspace{-4mm}
\subsection{Analysis of Iterative Rescoring}
We compare the non-iterative and iterative lattice rescoring methods using a Transformer LM. The parameter $\epsilon$ was set to 0.5 for the non-iterative method and 0.1 for the iterative one. The results of the iterative rescoring approach are in the last row in Table~\ref{tab:iterative}, and ``Score replacement'' refers to the refined operation of replacing $n$-gram LM scores on decoded lattices with neural LM ones. The 0.3\% absolute WER reduction on Hub5'00 from score replacement shows the benefit of neural LM over the original $n$-gram LM. The observation that score replacement performs worse than the non-iterative rescoring approach alone indicates the value of lattice expansion. 
\begin{table}[ht]
    \vspace{-2mm}
    \setlength{\tabcolsep}{2.0pt}
    \caption{WERs (\%) from proposed lattice rescoring strategies with a Transformer LM. }
    \label{tab:iterative}
    \centering
    \scalebox{0.9}{
    \begin{tabular}{ l  c  c  c }
    \toprule
        Rescoring Method & Hub5'00 & Swb & Callhm \\
        \midrule
        Non-iterative ($\epsilon$ = 0.5) & 10.6 & 6.8 & 14.3\\
        Score replacement & 10.8 & 6.8 & 14.6 \\
        Score replacement + Non-iterative & \textbf{10.3} & \textbf{6.6} & \textbf{14.0} \\
        \bottomrule
    \end{tabular}}
\end{table}
\vspace{-5mm}
\subsection{Comparison with $n$-gram Expansion}
\label{sec:expansion}
\begin{figure}[ht]
    \centering
    \begin{subfigure}[b]{0.35\textwidth}
        \centering
        \includegraphics[width=\textwidth]{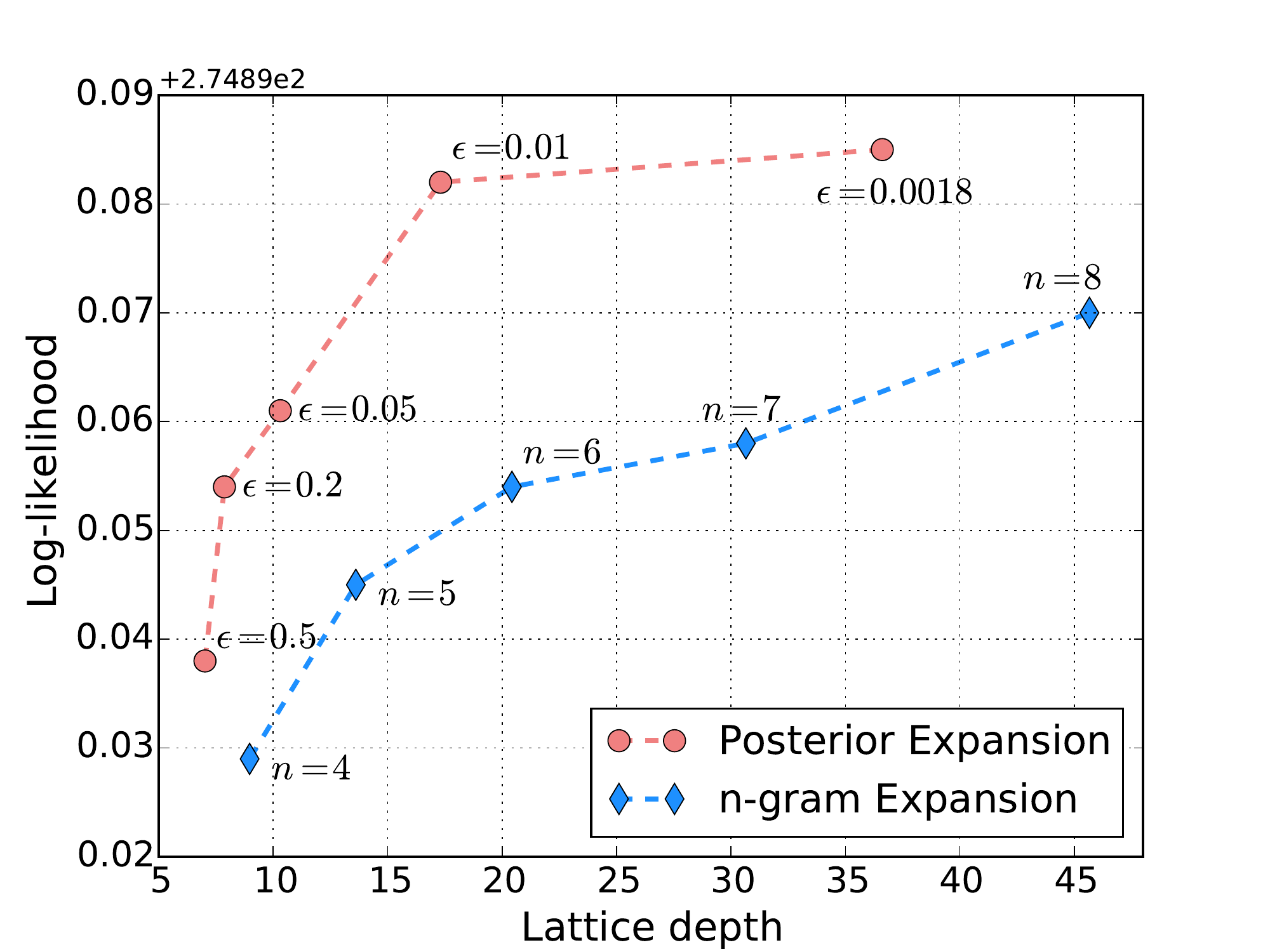}
        \caption{Log-likelihoods and lattice depths.}
        \label{fig:loglikelihoods}
    \end{subfigure}
    \hfill
    \begin{subfigure}[b]{0.35\textwidth}
        \centering
        \includegraphics[width=\textwidth]{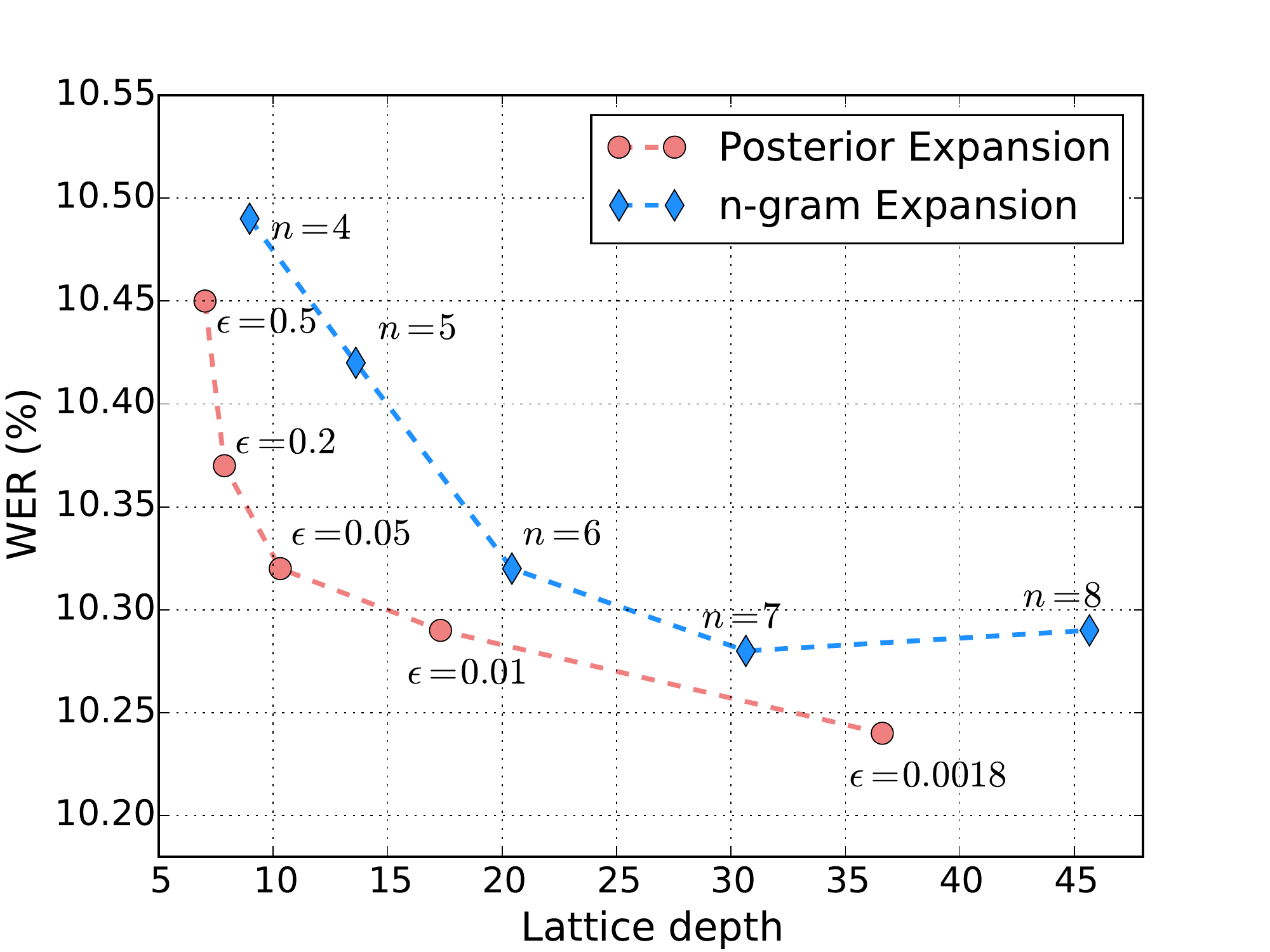}
        \caption{WERs and lattice depths.}
        \label{fig:expansion_comparison}
    \end{subfigure}
    \caption{Log-likelihoods, WERs, and lattice depths for different $\epsilon$ values and $n$-gram orders, respectively.}
    \vspace*{-\baselineskip}
\end{figure}

The performance of the proposed lattice expansion is compared with $n$-gram approximation based expansion using iterative rescoring strategy and a Transformer LM. The same pruning beam is used for a fair comparison. We evaluate the two methods through average log-likelihood of best paths and WER from rescored lattices in the Hub5'00 test set. Fig.~\ref{fig:loglikelihoods} summarizes the log-likelihoods and lattice depths measured in frame-level average number of arcs for different $\epsilon$ values and $n$-gram orders. We can observe that the posterior-based expansion method results in higher log-likelihoods than $n$-gram expansion. The corresponding WERs in Fig.~\ref{fig:expansion_comparison} show that the proposed expansion method generates more compact lattices with better recognition performance. Though WER is a more noisy metric, it essentially reflects the tendency of the log-likelihood curve. 
We can infer from the results that lattice rescoring with the new expansion method can be faster for even better recognition performance than $n$-gram expansion.

\vspace{-2mm}
\subsection{Comparison with Pruned Lattice Rescoring}
We compare the proposed non-iterative lattice rescoring method with the pruned lattice rescoring algorithm~\cite{xu2018pruned} in Kaldi. An LSTM LM trained with Kaldi RNNLM toolkit is used for experiments. WERs and lattice depths (measured as average number of arcs cross a frame on rescored lattices of Hub5'00 full set) are present in Table~\ref{tab:pruned_vs_ours}. The WERs from using a Kneser–Ney (KN) smoothed 4-gram LM and $N$-best rescoring are shown for reference.  
\begin{table}[ht]
    \vspace{-2mm}
    \setlength{\tabcolsep}{2.0pt}
    \caption{WERs (\%) and lattice depths from pruned lattice rescoring and the proposed non-iterative lattice rescoring.} 
    \label{tab:pruned_vs_ours}
    \centering
    \scalebox{0.9}{
    \begin{tabular}{ l c  c  c  c}
    \toprule
        \multirow{2}{*}{Method} & \multicolumn{3}{c}{WER} & 
        \multirow{2}{*}{Lattice Depth}\\
        \cmidrule {2-4} 
        & Hub5'00 & Swb & Callhm & \\
        \midrule
        4-gram KN & 12.8 & 8.6 & 17.0 & 31.5 \\
        $N$-best & 11.3 & 7.5& 15.0 &- \\
        Pruned (4-gram approx.) & 11.2 & \textbf{7.3} & 15.0 & 15.1  \\
        \midrule
        Non-iterative ($\epsilon$ = 0.5) & \textbf{11.1} & 7.4 & \textbf{14.9} & \textbf{6.4}\\
        \bottomrule
    \end{tabular}}
\end{table}

For both lattice rescoring methods, the interpolation weight of the LSTM LM (with the 4-gram LM) is 0.8. Compared with the pruned lattice rescoring, the non-iterative rescoring strategy obtains competitive performance and generates smaller lattices.

\subsection{Speedup}

In the proposed lattice rescoring strategies, the main speedups are from beam pruning and parallel computation of neural LM scores. Beam pruning reduces size of lattices by 3-4 times without degrading WERs in our experiments. The speedup by the batch computation varies with batch size and models. Compared with sequential evaluation, batch computation gives around 5-6 speedup with the PyTorch LSTM in a non-iterative rescoring setup. 

Compared with $N$-best rescoring (with different $N$s) in a parallel computation mode within each lattice as well, the proposed non-iterative lattice rescoring method accelerates the process by 1-3 times while obtains the same WERs. Besides, lattice rescoring with the Transformer LM is faster than with the PyTorch LSTM LM. This is expected considering the non-recurrent structure and fewer total parameters of the Transformer LM.

\subsection{WERs on SWBD}
We present WERs on SWBD with both $N$-best and lattice rescoring in Table~\ref{tab:wers}. The Transformer LM was used in both non-iterative and iterative rescoring methods. 
For $N$-best rescoring with the PyTorch trained LSTM, the state-carry trick~\cite{irie2019training} was used. 
$N$-best rescoring with the Transformer LM obtains slightly better performance than the PyTorch LSTM, consistent with the results from lattice rescoring in Table~\ref{tab:arc_estimation}. As expected, both non-iterative and iterative lattice rescoring methods obtain better recognition performance than $N$-best rescoring, and smaller expansion threshold generally leads to better WER.
However, since the computation cost for the iterative rescoring method is roughly doubled, non-iterative rescoring is more practical considering latency.
\vspace{-1mm}
\begin{table}[ht]
    \setlength{\tabcolsep}{2.0pt}
    \caption{WERs (\%) from proposed lattice rescoring strategies with a Transformer LM. $N$ is set to 20 for $N$-best rescoring.}
    \label{tab:wers}
    \centering
    \scalebox{0.9}{
    \begin{tabular}{ l  c  c  c }
    \toprule
        Method & Hub5'00 & Swb & Callhm \\
        \midrule
         4-gram KN & 12.8 & 8.6 & 17.0\\
        $N$-best (LSTM) & 10.9 & 7.1& 14.6 \\
        $N$-best (Transformer) & 10.8 & 7.2& 14.4  \\
        \midrule
        Non-iterative ($\epsilon$ = 0.5) & 10.6 & 6.8 & 14.3\\
        Non-iterative ($\epsilon$ = 0.005) & 10.4 & 6.8 & 14.0\\
        \midrule
        Iterative ($\epsilon$ = 0.1) & 10.3 & 6.6 & 14.0 \\
        Iterative ($\epsilon$ = 0.001) & \textbf{10.2} & \textbf{6.5}& \textbf{13.9} \\
        \bottomrule
    \end{tabular}}
\end{table}
\vspace{-3mm}
\section{Conclusion and Future Work}
In this work, we propose an efficient lattice rescoring strategy that computing neural LM scores within a lattice in parallel. The proposed method mainly consists of a posterior-based lattice expansion algorithm and a constrained path cover method for converting a lattice into a list representation. We also propose a refined rescoring strategy for further accuracy improvement.
Experiments on SWBD show that the posterior-based lattice expansion outperforms $n$-gram expansion. 
The proposed rescoring strategy obtains comparable WERs with marginally faster speed compared with the pruned lattice rescoring. To achieve the same recognition performance, the proposed rescoring method generally is faster than $N$-best rescoring in batch computation mode as well. 

The proposed parallel rescoring strategy makes it easier and more flexible to perform lattice rescoring with PyTorch-trained neural LMs in Kaldi. In the future, we plan to explore more effective ways of lattice expansion for further speedup. 

\vfill\pagebreak

\bibliographystyle{IEEEbib}
\bibliography{strings,refs}
\end{document}